\documentclass{article}
\usepackage{graphicx,amsmath,amsfonts,latexsym}
\newcommand{\var}{{\rm Var}}

\newcommand{\mse}{{\rm MSE}}

\newcommand{\eps}{\varepsilon}

\usepackage{graphicx}
\usepackage[round]{natbib}
\usepackage{epsfig}
\title{Predicting landfalling hurricane numbers from sea surface temperature: theoretical
comparisons of direct and indirect approaches}

\author{
Stephen Jewson (RMS)\footnote{\emph{Correspondence email}: \texttt{stephen.jewson@rms.com}}\\
Thomas Laepple (AWI)\\
Kechi Nzerem (RMS)\\
Jeremy Penzer (LSE)
}

\begin{document}
\maketitle

\begin{abstract}
We consider two ways that one might convert a prediction of sea
surface temperature (SST) into a prediction of landfalling
hurricane numbers. First, one might regress historical numbers of
landfalling hurricanes onto historical SSTs, and use the fitted
regression relation to predict future landfalling hurricane
numbers given predicted SSTs. We call this the direct approach.
Second, one might regress \emph{basin} hurricane numbers onto
historical SSTs, estimate the proportion of basin hurricanes that
make landfall, and use the fitted regression relation and
estimated proportion to predict future landfalling hurricane
numbers. We call this the \emph{indirect} approach. Which of these
two methods is likely to work better? We answer this question for two simple models.
The first model is reasonably realistic, but we have to resort to using
simulations to answer the question in the context of this model.
The second model is less realistic, but allows us to derive
a general analytical result.
\end{abstract}

\section{Introduction}

There is a great need to predict the distribution of the number of
hurricanes that might make landfall in the US in the next few
years. Such predictions are of use to all the entities that are
affected by hurricanes, ranging from local and national
governments to insurance and reinsurance companies. How, then,
should we make such predictions? There is no obvious best method.
For instance, one might consider making a prediction based on time-series analysis of the time-series of
historical landfalling hurricane numbers;
one might consider making a prediction of basin hurricane numbers using time-series analysis, and convert
that prediction to a prediction of landfalling hurricane numbers;
one might consider trying to predict SSTs first, and
convert that prediction to a prediction of landfalling numbers; or
one might try and use output from a numerical model of the climate
system. All of these are valid approaches, and each has their own
pros and cons.

In this article, we consider the idea of first predicting SST and then predicting
hurricane numbers given a prediction of SST. There are two obvious flavours of
this. The first is what we will call the `direct' (or `one-step') method, in which one regresses historical
numbers of landfalling hurricanes directly onto historical SSTs, and uses the
fitted regression relation to convert a prediction of future SSTs into a prediction
of future hurricane numbers. The second is what
we will call the `indirect' (or `two-step') method, in which one regresses \emph{basin} hurricane numbers onto
historical SSTs, predicts basin numbers, and then predicts landfalling numbers
from basin numbers. In the simplest version of the indirect method one might predict landfalling numbers
as a constant proportion of the number of basin hurricanes, where this proportion is
estimated using historical data.

Consideration of the direct and indirect SST-based methods motivates the question: at a theoretical level, which
of these two methods is likely to work best? This is a statistical question about the properties
of regression and proportion models. We consider this abstract question in the
context of two simple models.
The first model is the more realistic of the two.
It uses observed SSTs, models the mean number of hurricanes in the basin as a linear function of SST,
and models each basin hurricane as having a constant probability of making landfall.
We run simulations that allow us to directly compare the performance of the direct and indirect methods
in the context of this model.
The second model is less realistic, but allows us to derive a general analytical result for the relative
performance of the direct and indirect methods. In this model we represent SST, basin and landfalling hurricane numbers
as being normally distributed and linearly related.

We don't think the answer as to which of the direct or indirect methods is better is \emph{a priori} obvious.
On the one hand, the direct method has fewer parameters to estimate, which might work in its favour.
On the other hand, the indirect method allows us to use more data by incorporating the basin hurricane numbers
into the analysis.

Section~\ref{methods1} describes the methods used in the simulation study, and
section~\ref{results1} describes the results from that study.
In section~\ref{linearnormalmodel} we derive general analytic results for the linear-normal model.
Finally in section~\ref{summary} we discuss our results.

\section{Simulation-based analysis: methods}\label{methods1}

For our simulation study, we compare the direct and indirect methods described above as follows.

\subsection{Generating artificial basin hurricane numbers}

First, we simulate 10,000 sets of artificial basin hurricane numbers for the period 1950-2005,
giving a total of 10,000 x 56 = 560,000 years of simulated hurricane numbers.
These numbers are created by sampling from poisson distributions with mean given by:
\begin{equation}
\lambda=\alpha+\beta S
\end{equation}
where $S$ is the observed MDR SST for each year in the period 1950-2005.
The values of $\alpha$ and $\beta$ are derived from model 4 in table 7 in~\citet{e04a}, in which
observed basin hurricane numbers were regressed onto observed SSTs using data
for 1950-2005. They have
values of 6.25 and 5, respectively.

The basin hurricane numbers we create by this method should contain roughly the same long-term
SST driven variability as the observed basin hurricane numbers, but different numbers of
hurricanes in the individual years. We say `roughly' the same, because (a) the linear model
we are using to relate SST to hurricane numbers is undoubtedly not exactly correct, although given
the analysis in~\citet{e04a} is certainly seems to be reasonable, and (b) the
parameters of the linear model are only estimated.

\subsection{Generating artificial landfalling hurricane numbers}

Given the 10,000 sets of simulated basin hurricane numbers described above, we then
create 10,000 sets of simulated \emph{landfalling} hurricane numbers by applying the rule
that each basin hurricane has a probability of 0.254 of making landfall (this value is taken
from observed data for 1950-2005).

The landfalling hurricane numbers we create by this method should contain roughly the
same long-term SST driven variability as the observed landfalling series, but different
numbers of hurricane in the individual years. They should also contain roughly the right
dependency structure between the number of hurricanes in the basin and the number at landfall
(e.g. that years with more hurricanes in the basin will tend to have more hurricanes at landfall).

\subsection{Making predictions}

We now have 10,000 sets of 56 years of artificial data for basin and landfalling hurricanes.
This data contains a realistic representation of the SST-driven variability of hurricane numbers, and
of the dependency structure between the numbers of hurricanes in the basin and at landfall,
but different actual numbers of hurricanes from the observations. We can consider this data
as 10,000 realisations of what might have occurred over the last 56 years, had the SSTs
been the same, but the evolution of the atmosphere different. This data is a test-bed
that can help us understand aspects of the predictability of landfalling hurricanes given SST.

\begin{figure}[!hb]
  \begin{center}
    \rotatebox{-90}{\scalebox{0.7}{\includegraphics{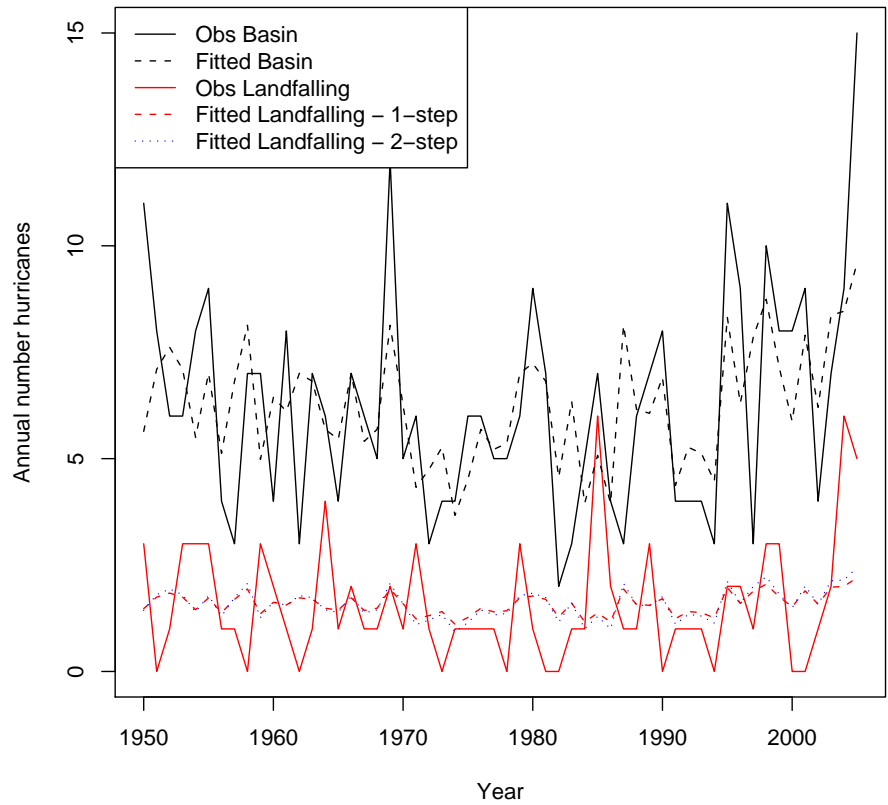}}}
  \end{center}
    \caption{
Atlantic basin and landfalling hurricane numbers for the period 1950 to 2005 (solid lines),
with the component of the variability that can be explained by SSTs (broken lines).
}
     \label{f01}
\end{figure}

\begin{figure}[!hb]
  \begin{center}
    \rotatebox{-90}{\scalebox{0.7}{\includegraphics{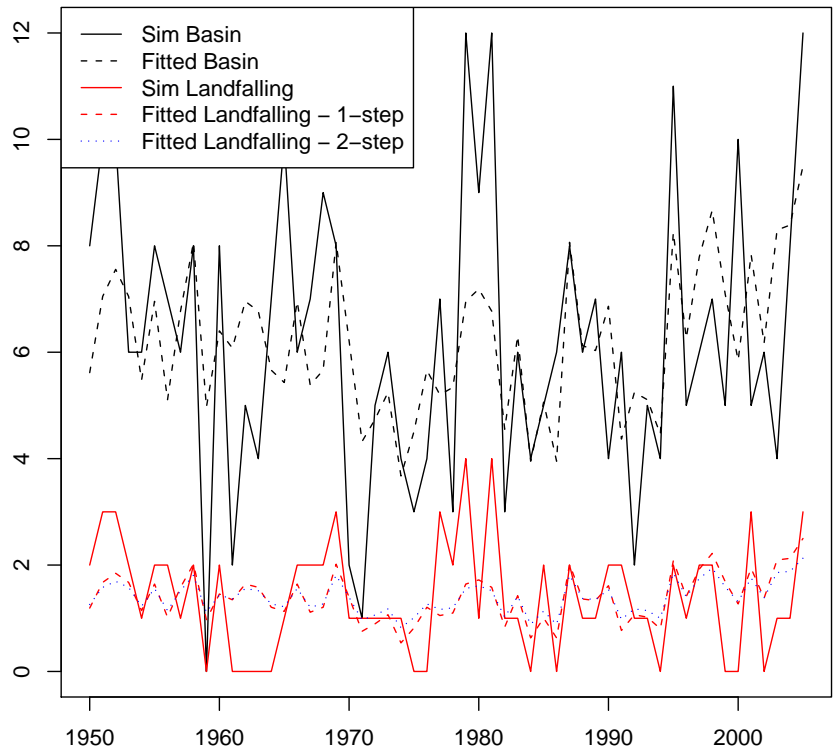}}}
  \end{center}
    \caption{
One realisation of simulated basin and landfalling hurricane numbers (solid lines),
with the SST driven components (broken lines).
}
     \label{f02}
\end{figure}

\begin{figure}[!hb]
  \begin{center}
    \rotatebox{-90}{\scalebox{0.7}{\includegraphics{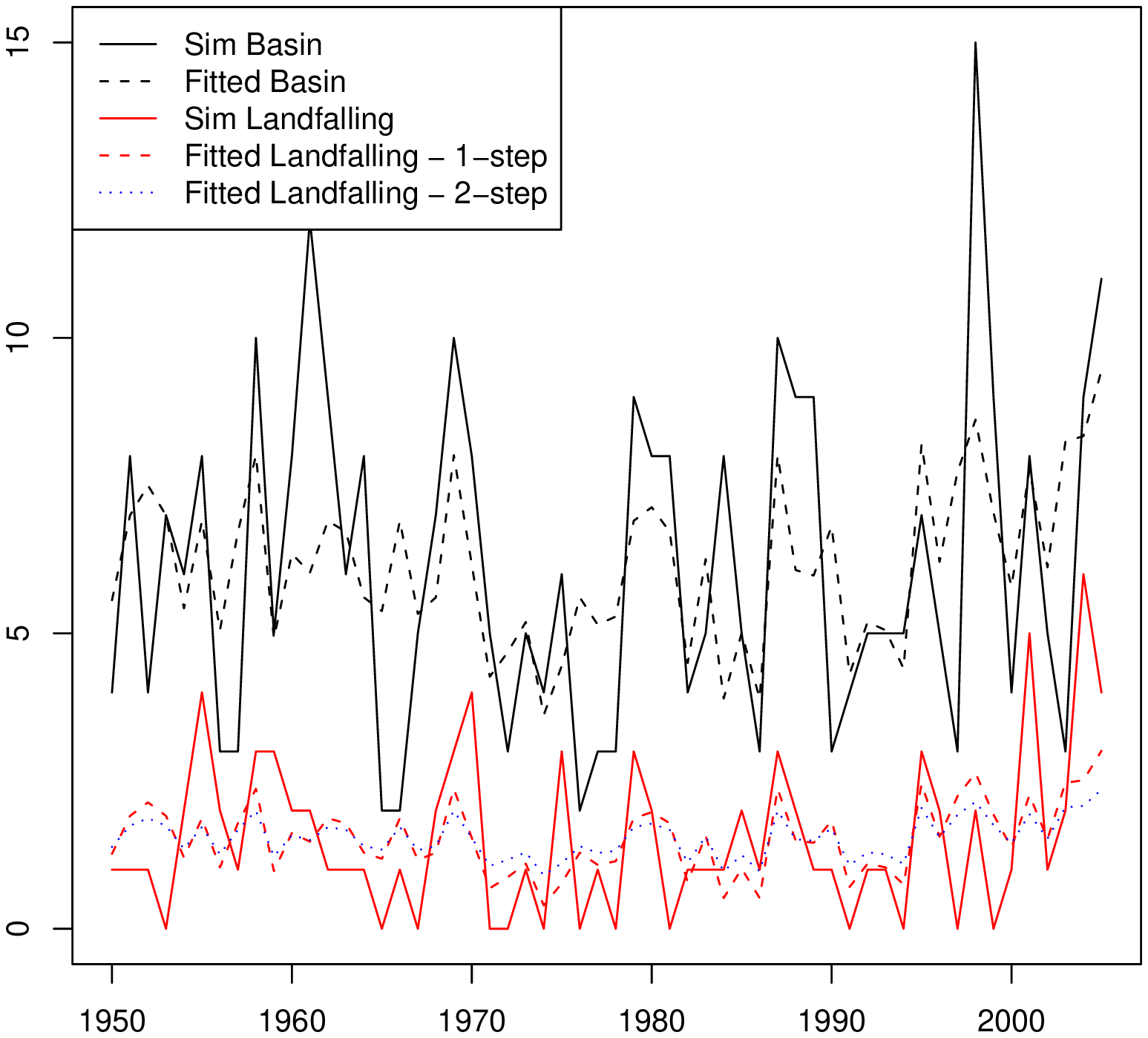}}}
  \end{center}
    \caption{
As in figure~\ref{f02}, but for a different realisation.
}
     \label{f03}
\end{figure}

\begin{figure}[!hb]
  \begin{center}
    \rotatebox{-90}{\scalebox{0.7}{\includegraphics{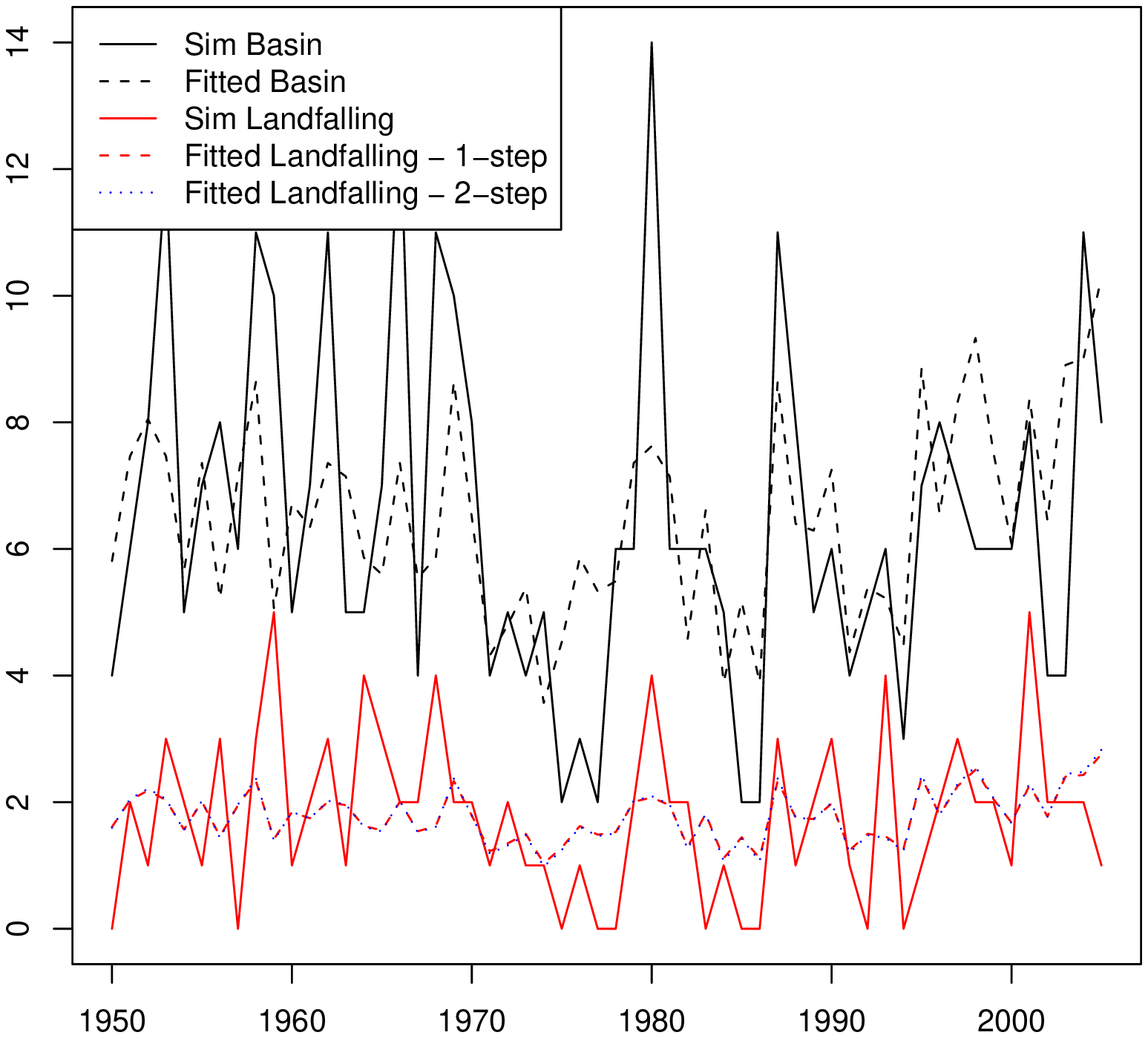}}}
  \end{center}
    \caption{
As in figure~\ref{f02}, but for a different realisation.
}
     \label{f04}
\end{figure}

\begin{figure}[!hb]
  \begin{center}
    \rotatebox{-90}{\scalebox{0.7}{\includegraphics{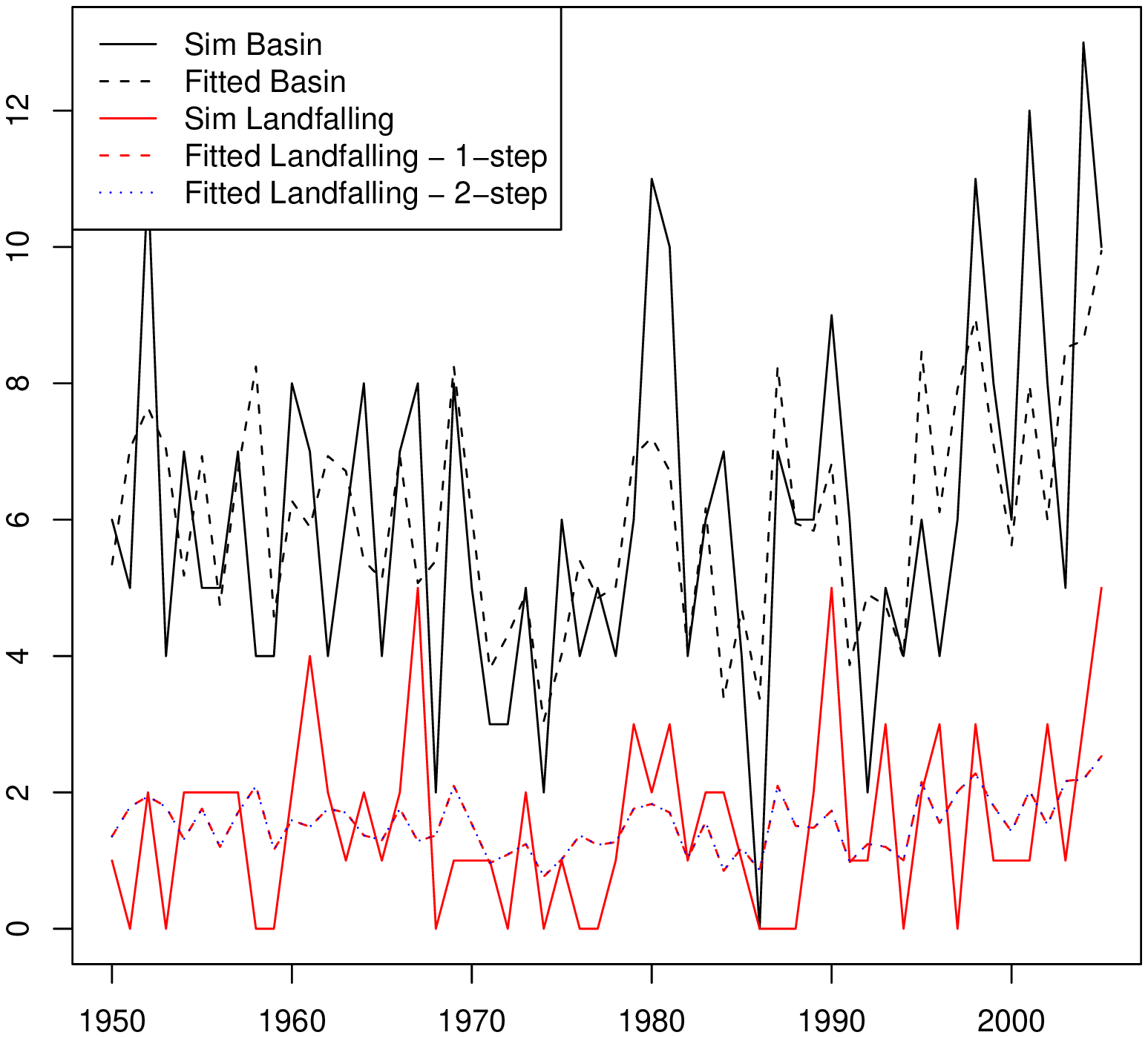}}}
  \end{center}
    \caption{
As in figure~\ref{f02}, but for a different realisation.
}
     \label{f05}
\end{figure}

The observed and simulated data is illustrated in figures~\ref{f01} to \ref{f05}.
Figure~\ref{f01} shows the observed basin data (solid black line) and the observed
landfall data (solid grey line). The dashed black line shows the variability in the
observed basin data that is explained using SSTs. The dotted grey line shows the variability
in the observed landfall data that is explained using SSTs using the direct method, and
the dotted grey line shows the variability in the landfall data that is explained using
SSTs using the indirect method.

Figures~\ref{f02} to \ref{f05} show 4 realisations of the simulated data. In each figure
the dotted and dashed lines are the same as in figure~\ref{f01}, and show the SST driven
signal. The solid black line then shows the simulated basin hurricane numbers and the solid
grey line shows the simulated landfalling hurricane numbers.

We test predictions of landfalling hurricane numbers using the direct method as follows:

\begin{itemize}

    \item we loop through the 10,000 sets of simulated landfalling hurricanes

    \item for each set, we miss out one of the 56 years

    \item using the other 55 years in that set, we build a linear regression model between
    SST and landfalling hurricane numbers

    \item we then use that fitted model to predict the number of landfalling hurricanes in
    the missed year, given the SST for that year

    \item we calculate the error for that prediction

    \item we then repeat for all 10,000 sets (missing out a different year each time)

    \item this gives us 10,000 prediction errors, from which we calculate the RMSE

\end{itemize}

We test the indirect method in almost exactly the same way, except that this time we
also fit a model for predicting landfalling numbers from basin numbers.

\subsection{Comparing the predictions}

We compare the direct and indirect predictions in two ways:

\begin{itemize}

    \item First, we compare the two RMSE values

    \item Second, we count what proportion of the time the errors from the direct method
    are smaller than the errors from the indirect method

\end{itemize}

We also repeat the entire calculation a number of times as a rough way to evaluate the
convergence of our results.

\section{Simulation-based analysis: results}\label{results1}

We now present the results from our simulation study.
The RMSE for the direct method is 1.61 hurricanes, while the RMSE for the indirect method is 1.58 hurricanes.
This difference is small, but the sign of it does appear to be real: when we repeat the whole experiment
a number of times, we always find that the indirect method beats the direct method.

The indirect method beats the direct method 51.8\% of the time.

Given the design of the experiment, these results tell us how the two methods perform,
on average over the whole range of SST values. Next year's SST, however, is likely to be warm
relative to historical SSTs. We therefore also consider the more specific question of how the methods
are likely to perform for given warm SSTs. Based on~\citet{e20}, we fit a linear trend to the historical
SSTs, and extrapolate this trend out to 2011. This then gives SST values that are warmer than
anything experienced in history (27.987$^o$C to be precise). We then repeat the whole analysis
for predictions for this warm SST only. The results are more or less as before: the indirect
method still wins, only this time by a slightly larger margin. The ratio of RMSE scores
(direct divided by indirect) increases from 1.02 to 1.04.

\section{The Linear normal case}\label{linearnormalmodel}

We now study a slightly less realistic model, in which we take SSTs and hurricane numbers
in the basin and at landfall to be normally distributed. These changes allow us to derive
a very general result for the relative performance of the direct and indirect methods.

\subsection{The setup}

Here's how we set the problem up in this case.

Consider two simple regression models for centred random
variables $Y$ and $Z$,
\begin{eqnarray*}
Y &=& X \beta + \eps, \quad  \eps \sim (0,\sigma_\eps^2 I_n), \\
Z &=& Y \gamma + \eta, \quad  \eta \sim (0,\sigma_\eta^2 I_n),
\end{eqnarray*}

where $\eps$ and $\eta$ are independent. Here $X$, $Y$, $Z$,
$\eps$ and $\eta$ are $n \times 1$ column vectors, $\beta$ and
$\gamma$ are scalars, and $I_n$ is the $n \times n$ identity
matrix. We will assume $X$ is fixed.

In relation to the hurricane problem, $X$ is the time-series of $n$ years
of SST values, $Y$ is the time-series of $n$ years of basin hurricane numbers and $Z$ is
the time-series of $n$ years of landfalling hurricane numbers. Note that in our
notation $X$ is the \emph{whole time-series} of SST, written as a vector, and similarly
for $Y$ and $Z$. Using vector notation avoids the messy use of subscripts.
Two immediate comments about this setup: (a) we are assuming that basin and landfalling hurricane
numbers are normally distributed. This doesn't really make sense, since they are counts that
can only take integer values: using a poisson distribution would make more sense.
We are starting off by addressing this question for normally distributed data
because it's more tractable that way;
(b) we are assuming a linear relationship (with offset and slope) between basin hurricanes
and landfalling hurricanes. This is also a little odd, since there is no reason to have an offset
in this relationship: if there aren't any basin hurricanes, there can't be any landfalling hurricanes.
The most obvious model would be that each hurricane has a constant proportion of making landfall.
Again, we are starting off by addressing this question in a linear context because it's more
tractable that way.

We want to know about the accuracy of forecasts that we might make with the direct and
indirect methods. This translates mathematically into saying that we want to estimate
\begin{eqnarray}
E(z_{n+1}) &=& E(y_{n+1}) \gamma\\
           &=& x_{n+1} \beta \gamma\\
           &=& x_{n+1}\delta
\end{eqnarray}

where $\delta = \beta \gamma$.

The problem then boils down to measuring
the quality of the estimator of $\delta$ since, if $\hat{z}_{n+1}
= x_{n+1} \hat{\delta}$ is an estimator of $E(z_{n+1})$ then
\begin{eqnarray}
\mse(\hat{z}_{n+1}) &=& \mse(x_{n+1} \hat{\delta})\\
                    &=& E[(x_{n+1}\hat{\delta} - x_{n+1} \delta)(x_{n+1} \hat{\delta} - x_{n+1}\delta)']\\
                    &=& x_{n+1} \mse(\hat{\delta}) x_{n+1}'.\label{mse}
\end{eqnarray}

So we now consider the direct and indirect methods for estimating $\delta$.

\subsection{Direct estimator of $\delta$}

We start by considering the direct, or one-step, method.
This means we consider the relationship between $X$ and $Z$, ignoring $Y$. The
usual OLS estimator for $\delta$ is
\begin{eqnarray}
\delta^\dagger &=& (X'X)^{-1} X' Z\\
               &=& (X'X)^{-1} X'(X \beta \gamma +\eps \gamma + \eta)\\
               &=& \delta + (X'X)^{-1} X' (\eps \gamma + \eta).
\end{eqnarray}

What are the statistical properties of this estimator?

In terms of mean:
\begin{equation}
E(\delta^\dagger) = \delta
\end{equation}
i.e. the estimator is unbiased.

In terms of variance
\begin{eqnarray}
\var(\delta^\dagger) &=& (X'X)^{-1} X' \var(\eps \gamma + \eta) X(X'X)^{-1}.
\end{eqnarray}

We know that $\var(\eps \gamma + \eta) = \sigma_\eps^2 I_n
\gamma^2 + \sigma_\eta^2 I_n$, so
\begin{equation}\label{vard1}
\var(\delta^\dagger) = (X'X)^{-1} (\sigma_\eps^2 \gamma^2 + \sigma_\eta^2).
\end{equation}

By equation~\ref{mse} this then gives us an expression for the performance of the direct method.

\subsection{Indirect estimator of $\delta$}

We now consider the indirect, or two-step, method.
This means considering the relationships between $X$ and $Y$, and $Y$ and $Z$.

First, we consider estimating each regression separately. The OLS estimators for the slopes in each case are:
\begin{eqnarray}
\hat{\beta} &=& (X'X)^{-1} X' Y \\
            &=& \beta + (X'X)^{-1} X' \eps \\
\hat{\gamma} &=& (Y'Y)^{-1} Y' Z\\
             &=& \gamma + (Y'Y)^{-1} Y' \eta
\end{eqnarray}

We now put the two models together, to create a single regression model based on the separate estimates
for the two steps. We call the estimate of the slope of this combined model $\hat{\delta}$.
Combining the expressions above, we have that:
\begin{eqnarray}
\hat{\delta} &=& \hat{\beta} \hat{\gamma}\\
             &=& \beta \gamma +(X'X)^{-1}X' \eps \gamma + \beta (Y'Y)^{-1} Y' \eta + (X'X)^{-1} X' \eps (Y'Y)^{-1} Y' \eta
\end{eqnarray}

What are the statistical properties of this estimator $\hat{\delta}$?

It is clear (by independence of $\eps$ and $\eta$) that
$\hat{\delta}$ is unbiased;
\begin{eqnarray}
E(\hat{\delta}) &=& \beta \gamma\\
                &=&\delta
\end{eqnarray}

The variance is more awkward. Note that if $\eps$ were known then
$\hat{\beta}$ and $Y$ would be fixed constants. Thus,
\begin{eqnarray}
E(\hat{\delta}| \eps) &=&E(\hat{\beta} \hat{\gamma}| \eps)\\
                      &=& \hat{\beta} E(\hat{\gamma}|\eps)\\
                      &=& \hat{\beta} \gamma, \\
\var(\hat{\delta}| \eps) &=& \var(\hat{\beta} \hat{\gamma}|\eps)\\
                         &=& \hat{\beta} \var(\hat{\gamma}|\eps) \hat{\beta}'\\
                         &=& \hat{\beta} (Y'Y)^{-1} \hat{\beta}' \sigma_\eta^2.
\end{eqnarray}

and so
\begin{eqnarray}
\var(\hat{\delta}) &=& \var(\hat{\beta} \hat{\gamma})\\
                   &=& E[\var(\hat{\beta} \hat{\gamma}|\eps)] + \var[E(\hat{\beta} \hat{\gamma}| \eps)]\\
                   &=& E[\hat{\beta} (Y'Y)^{-1} \hat{\beta}'] \sigma_\eta^2 + \gamma \var(\hat{\beta}) \gamma'.
\end{eqnarray}

where we have used a standard relation for disaggregating the variance:
\begin{equation}
\mbox{var}(a)=E[\mbox{var}(a|b)]+\mbox{var}[E(a|b)]
\end{equation}

Using the facts that
\begin{eqnarray}
E(Y'Y) & = &  \beta' X' X \beta + n \sigma_\eps^2 \\
E(\hat{\beta} \hat{\beta}') &=& \beta \beta' + (X'X)^{-1}
\sigma_\eps^2
\end{eqnarray}

and approximating to second order:

\begin{equation}\label{vard2}
\var(\hat{\delta}) =
\left[\frac{\beta^2 + q^2}{\beta^2  + n q^2}\right] (X'X)^{-1}
\sigma_\eta^2 + q^2 \gamma^2.
\end{equation}

where $q^2=(X'X)^{-1} \sigma_\eps^2$.

\subsection{Comparing the two estimators}

We are now in a position to compare the estimators for the direct and indirect methods.
Subtracting equation~\ref{vard2} from equation~\ref{vard1} gives:

\begin{eqnarray}
\var(\delta^\dagger)-\var(\hat{\delta})
 &=&(X'X)^{-1} (\sigma_\eps^2 \gamma^2 + \sigma_\eta^2)
    -\left[\frac{\beta^2 + q^2}{\beta^2  + n q^2}\right] (X'X)^{-1}\sigma_\eta^2 - (X'X)^{-1} \sigma_\eps^2 \gamma^2\\
 &=&(X'X)^{-1} \sigma_\eta^2
    -\left[\frac{\beta^2 + q^2}{\beta^2  + n q^2}\right] (X'X)^{-1}\sigma_\eta^2\\
 &=&\left(1-\left[\frac{\beta^2 + q^2}{\beta^2  + n q^2}\right]\right) (X'X)^{-1}\sigma_\eta^2\\
 &=&\left[\frac{(n-1) q^2}{\beta^2 + n q^2}\right] (X'X)^{-1} \sigma_\eta^2
\end{eqnarray}

The right hand side of this equation is clearly positive for $n>1$.

This indicates:
\begin{itemize}

\item that using the indirect method is an improvement on the direct method,
at least up to our second order approximations

\item that if $\frac{\beta^2}{q^2}$ is small
or $\sigma_\eta^2$ large then using the indirect method
provides a marked improvement over the direct approach
\end{itemize}

\section{Conclusions}\label{summary}

We have compared the likely performance of direct and indirect methods for predicting landfalling hurricane numbers
from SST. The direct method is based on building a linear regression model directly
from SST to landfalling hurricane numbers. The indirect method is based on building a regression model
from SST to basin numbers, and then predicting landfalling numbers from basin numbers using
a constant proportion.

First, we compare these two methods in the context of a reasonably realistic model, using simulations.
We find that the indirect method is better than
the direct method, but that the difference is small.

Secondly, we compare the two methods in the context of a less realistic model in which
all variables are normally distributed. For this model
we are able to derive the interesting general result that the indirect method should \emph{always} be better.

Which method should we then use in practice?
If we had to chose one method, our results seem to imply that we should choose the indirect method,
since it is more accurate.
The simulation results suggest, however, that the performance of the two methods is likely to be very close
for the values of the parameters appropriate for hurricanes in the real world.
Given the possibility to use two methods we would use both, as alterative points of view.

Ideally we would also be able to solve the more realistic model analytically, as we have done for the
linear-normal case. We are working on that.

\bibliography{arxiv}

\newpage

\newpage

\newpage

\newpage

\newpage

\end{document}